\documentclass[11pt]{article}
\usepackage{graphicx}
\usepackage{amsmath}
\usepackage{amsfonts}
\usepackage{mathtools}
\usepackage{color}
\usepackage{amssymb}
\usepackage{cite}
\usepackage{dcolumn}
\usepackage{bm}
\usepackage{microtype} 
\usepackage[linktoc=all]{hyperref}
\usepackage[capitalize]{cleveref}
\usepackage{braket}
\def\mean#1{\left< #1 \right>}
\numberwithin{equation}{section}
\newcommand{\lag}{\mathcal{L}}
\newcommand{\ud}{\mathrm{d}}

\title{}
\author{}

\numberwithin{equation}{section}

\begin{document}
	%
	\renewcommand{\thefootnote}{\fnsymbol{footnote}}
	~
	\vspace{2.5truecm}
	\begin{center}
		{\LARGE \bf{Field Theories and Fluids for an Interacting Dark Sector}}
	\end{center} 
	\vspace{1truecm}
	\thispagestyle{empty} \centerline{
		{\Large  {Mariana Carrillo Gonz\'alez,}}
		{\Large and} 
		{\Large Mark Trodden}
	}
	
	\vspace{.5cm}
	
	\centerline{{\it Center for Particle Cosmology, Department of Physics and Astronomy,}}
	\centerline{{\it University of Pennsylvania, Philadelphia, PA 19104, USA}}

	\vspace{.5cm}
\begin{abstract}
\noindent
We consider the relationship between fluid models of an interacting dark sector, and the field theoretical models that underlie such descriptions. This question is particularly important in light of suggestions that such interactions may help alleviate a number of current tensions between different cosmological datasets. We construct consistent field theory models for an interacting dark sector that behave exactly like the coupled fluid ones, even at the level of linear perturbations,  and can be trusted deep in the nonlinear regime. As a specific example, we focus on the case of a Dirac, Born-Infeld (DBI) field conformally coupled to a quintessence field. We show that the fluid linear regime breaks before the field gradients become large; this means that the field theory is valid inside a large region of the fluid nonlinear regime.
\end{abstract}

\newpage

\setcounter{tocdepth}{2}
\newpage
\renewcommand*{\thefootnote}{\arabic{footnote}}
\setcounter{footnote}{0}

\section{Introduction: Motivations for an Interacting Dark Sector}
The standard cosmological model has proven to be successful at explaining most observations of our universe. Despite this success, several anomalies and tensions have been found between cosmological and astrophysical data. These may point to the presence of new physics, although it is important to note that comparing these datasets is a nontrivial task. The estimates found from the cosmic microwave background (CMB) data, for example, are model dependent; on the other hand, most astrophysical sources give direct estimates, but for which the backgrounds are largely complex. One much-discussed discrepancy is between measurements of the Hubble parameter at different redshifts. The local measurement obtained by the observation of Cepheid variables gives $H_0=73.24\pm1.79\,\text{km}/(\text{s Mpc})$ \cite{Riess:2016jrr}, which is 3.4 sigma higher than the Planck data estimate of $H_0=66.93\pm-0.62\,\text{km}/(\text{s Mpc})$ \cite{Ade:2015xua}, found by assuming a $\Lambda$CDM model with 3 neutrinos with masses of $0.06$ eV. It has been suggested that this discrepancy might be alleviated by introducing an interaction between dark matter and dark energy \cite{DiValentino:2017iww,Salvatelli:2013wra,Pettorino:2013oxa,Pettorino:2012ts}. Such interacting dark sector (IDS) models have been considered before, and allow for an energy transfer between the dark matter and dark energy caused by an interaction between the two sectors. The presence of this interaction gives rise to an expansion history slightly different from the $\Lambda$CDM one, and modifies the growth of structure.
 
The phenomenological approach to dark energy coupled to dark matter treats both components as perfect fluids. In these coupled models, the stress-energy tensors of dark matter and dark energy, instead of being conserved independently, satisfy
\begin{equation}
\nabla_\mu T^{\mu\nu}_\text{cdm}=-\nabla_\mu T^{\mu\nu}_\text{de}=Q^\nu=\xi H u^\nu \rho_\text{cdm/de} \ ,
\label{coupledemtensors}
\end{equation}
where $\rho_\text{cdm/de}$ either stands for $\rho_\text{cdm}$, the dark matter density, or $\rho_\text{de}$, the dark energy density; $H$ is the Hubble parameter; and $\xi$ is usually taken to be a constant, although more fundamental field theory models can give rise to a non-constant $\xi$. Treating dark matter and dark energy as fluids, the above equations can be written as
\begin{eqnarray}
\dot{\rho}_{cdm}+3H\rho_{cdm}=Q \label{Qcdm}\\ 
\dot{\rho}_{de}+3H(1+w_{de})\rho_{de}=-Q \label{Qde}\ ,
\end{eqnarray}
where $Q=\xi H \rho_\text{cdm/de}$. If $Q>0$, energy is transferred from dark energy to dark matter, and if $Q<0$, the situation is reversed. 

Models of dark matter treated as a perfect fluid interacting with a quintessence field have been extensively studied; see for example \cite{Amendola:1999er,Bean:2008ac,Koivisto:2005nr,Amendola:2003eq,Amendola:2003wa,Amendola:2002bs,Bean:2007nx,Bean:2007ny, LaVacca:2009yp,Pettorino:2008ez}. In these models, the dark matter energy density dilutes faster than $a^{-3}$. In \cite{Pettorino:2013oxa,Pettorino:2012ts} an interacting model with an inverse power law potential and coupling $\xi=\frac{\alpha'(\phi)\dot\phi}{H}$, where $\alpha\equiv\beta \phi/M_{Pl}$, was analyzed. It was found that a mean value of  $\beta\sim-0.066$\footnote{Notice that our definition of $\beta$ differs by a minus sign from the definition in \cite{Pettorino:2013oxa}.}, which differs from zero at $3.6\,\sigma$, can alleviate the $H_0$ tension between the Planck and Hubble Space Telescope (HST) datasets. These results are obtained by considering the linear regime of the theory. However, a coupling in the dark sector can effectively reduce or increase, depending on its sign, the friction term in the overdensity equation, and can also cause dark matter to feel an augmented Newtonian potential. Both of these effects can lead to important changes in the nonlinear regime. This has been explored in \cite{Wintergerst:2010ui,Mainini:2006zj,Baldi:2010td,Baldi:2010vv,Saracco:2009df,Casas:2015qpa}, where it was shown that even a small coupling, resulting in small differences with respect to $\Lambda$CDM in the linear regime, could lead to more significant differences in the nonlinear one; for example, modifying the predictions for the number of clusters. Given these discrepancies when comparing to $\Lambda$CDM, it is appealing to have an underlying field theoretical description that is valid deep in the nonlinear regime.

In this paper, we focus on the relation between fluid and field theoretical models of an IDS. Firstly, we consider the constraints imposed by quantum corrections on the IDS field theory models in Section \ref{field}. In Section \ref{osc}, we briefly review the case of a coupled model in which dark matter is an axion-like particle and then, in Section \ref{pofx}, we extend previous work by constructing field theory models that behave like the fluid models deep in the nonlinear regime. These models consist of a $P(X)$ field conformally coupled to the dark energy field. We find that the field gradients become large only when we are deep in the nonlinear regime. Once the gradients are large, caustics can form and the effective field theory is not valid. Finally, we conclude in Section \ref{conclu}.

\section{Field Theory Models of IDS} \label{field}
Since our goal here is to understand the relationship between the fluid description and the underlying field theoretical description, we now turn to field theory models in which the dark energy interaction is parametrized by a dimensionless function $\alpha(\phi)$. Models with similar features naturally arise in higher dimensional theories with branes, such as the Randall Sundrum I model \cite{Randall:1999ee}, and in Brans-Dicke theory after a conformal transformation \cite{Amendola:1999qq,Wetterich:1994bg}. We consider the general action
\begin{align}
S&=\int\ud^4x\sqrt{-g}\left[\frac{1}{2}M_{Pl}^2R-\frac{1}{2}\left(\nabla\phi\right)^2-V(\phi)\right] +S_{\chi} \left[e^{2\alpha(\phi)} g_{\mu\nu},\chi\right]+\sum_jS_j\left[g_{\mu\nu},\psi_j\right] \ , \label{inter}
\end{align}
where $\chi$ is the dark matter and $\psi_j$ are the standard model fields.
In order to ensure we can trust the effective field theory in the regime we wish to use it, we will examine the behavior of the model under quantum corrections before analyzing the evolution of the interacting model in more detail.

\subsection{Quantum corrections} \label{qc}
The aim here is to analyze the constraints imposed by quantum corrections to the coupling strength and the dark matter mass. To begin, consider the case of scalar dark matter. In this case, there is an interaction term
\begin{equation}
e^{\alpha(\phi)}V(\chi)\supset e^{\alpha(\phi)}m_\chi^2\chi\chi \ .
\end{equation}
If $\alpha(\phi)\equiv\beta\phi/M_{\text{Pl}}\ll1$, we may expand this term as
\begin{equation}
m_\chi^2\chi\chi+\frac{\beta}{M_{\text{Pl}}} m_\chi^2 \phi \chi\chi+\cdots\ ,
\label{expandedalpha}
\end{equation}
and the second term in this expression can lead to strong constraints when treating the full quantum theory. In fact, the leading order correction to the dark energy mass from this term yields, 
\begin{equation}
\Delta m_\phi\propto \beta \frac{m_\chi^2}{M_{\text{Pl}}} \ .
\end{equation}
In order for the field $\phi$ to behave as dark energy we need $m_\phi\lesssim H_0$. Thus, for the dark energy mass to remain technically natural, we require this to be a subdominant correction to the bare mass, i.e. that $\Delta m_\phi\ll m_\phi\sim H_0$. This leads to an upper limit on the dark matter mass:
\begin{equation}
m_\chi \ll 10^{-2} \text{eV} \sqrt{\frac{10^{-2}}{\beta}} \ ,
\end{equation}
where we have assumed that $\langle\phi\rangle\sim M_{\text{Pl}}$ during dark energy domination. As one can see from this bound, this points to a dark matter candidate which is an ultra-light boson that was never in equilibrium with the thermal bath. In order for it to have the required properties of dark matter, we also need its mass to be larger than the Hubble rate and, in fact, if $\chi$ is to constitute all the observed dark matter, even tighter constraints apply. In \cite{Hlozek:2014lca}, it was shown, using data from the Planck satellite, that in order for an ultra-light boson to account for all dark matter, its mass must be larger than $10^{-33}\text{GeV}$. With this constraint satisfied, such an uncoupled field would be indistinguishable from cold dark matter (CDM). Putting all these constraints together gives the allowed mass range for a coupled ultra-light boson constituting all dark matter, namely
\begin{equation}
10^{-24}\text{eV} \ll m_\chi \ll 10^{-2} \text{eV} \sqrt{\frac{10^{-2}}{\beta}} \ .
\end{equation}
The analogous analysis can also be carried out for fermionic dark matter~\cite{D'Amico:2016kqm}, and shows that an extremely small coupling is required to keep the quantum corrections small.  Assuming, for example, a standard WIMP thermal relic with $m_\chi\geq10$ GeV, this constraint implies that $\beta\lesssim 10^{-26}$.

In order to have a consistent model with a coupling that can alleviate the $H_0$ tension, we will therefore focus on scalar fields as dark matter. One option that has already been explored, for example in \cite{D'Amico:2016kqm}, is a coupled axion like particle. In the following section we review this case. A second option, that we will consider in the balance of this paper, is to consider coupled scalar fields with non-canonical kinetic terms. These could arise naturally in a number of models of high energy physics and, as we will show, can behave exactly as the fluid models of a coupled dark sector.

\section{Coupled oscillating scalar field dark matter} \label{osc}
The simplest example of a viable field theory model of an IDS consists of a quintessence field coupled to an oscillating scalar field which behaves as dark matter. Defining $\tilde{g}_{\mu\nu}\equiv e^{2\alpha(\phi)}g_{\mu\nu}$, the action for this field takes the deceptively canonical form
\begin{equation}
S_{\chi} = \int\ud^4x\sqrt{-\tilde{g}}\left(-\frac{1}{2}\left(\tilde\nabla\chi\right)^2-U(\chi)\right) \ ,
\end{equation}
where $\tilde{\nabla}$ is the covariant derivative for the metric $\tilde{g}_{\mu\nu}$ .The resulting equations of motion for the full action~(\ref{inter}) are then
\begin{eqnarray}
M_{Pl}^{2}G_{ab} =T^\text{SM}_{ab}+\nabla_{a}\phi\nabla_{b}\phi &-& \frac{1}{2}g_{ab}(\nabla\phi)^{2}-V(\phi)g_{ab} \nonumber \\
&+& e^{2\alpha(\phi)}\left(\nabla_{a}\chi\nabla_{b}\chi-\frac{1}{2}g_{ab}(\nabla\chi)^{2}-e^{2\alpha(\phi)}U(\chi)g_{ab}\right) \ ,
\end{eqnarray}
\begin{eqnarray}
\nabla_{a}\nabla^{a}\phi-V'(\phi) &=&\alpha'(\phi)e^{2\alpha(\phi)}\left(-\left(\nabla\chi\right)^2-4e^{2\alpha(\phi)}U(\chi)\right) \ ,\\
\nabla_{a}\nabla^{a}\chi-e^{2\alpha(\phi)}U'(\chi)&=&-2\alpha'(\phi) \nabla_\mu \phi \nabla^\mu \chi \ .
\end{eqnarray}
Since we are assuming that the standard model sector is minimally coupled to the Einstein frame metric, the energy-momentum tensor of that sector is conserved independently. However, by construction, the energy-momentum tensors of the dark matter and dark energy are not independent, and instead are related via~\eqref{coupledemtensors}
\begin{equation*}
\nabla_\mu T^{\mu\nu}_\text{cdm}=-\nabla_\mu T^{\mu\nu}_\phi=Q^\nu \ ,
\end{equation*}
where we are defining $T^{\mu\nu}_\phi$ without the coupled terms as
\begin{equation}
T^{\mu\nu}_\phi\equiv\nabla_{a}\phi\nabla_{b}\phi-\frac{1}{2}g_{ab}(\nabla\phi)^{2}-V(\phi)g_{ab} \ ,
\end{equation}
and $Q^\nu$ is given by
\begin{equation}
Q^\nu = -\alpha'(\phi)e^{2\alpha(\phi)}\nabla^\nu\phi\left((\nabla\chi)^{2} +4e^{2\alpha(\phi)}U(\chi)\right) \ .
\end{equation}

To investigate the evolution of the background cosmology in this model, we specialize to the flat FRW metric $ds^2=-dt^2 +a(t)^2 d{\bf x}^2$, so that the equations of motion now read
\begin{eqnarray}
3M_{\mathrm{p}}^{2}\displaystyle H^{2} &=&\frac{1}{2}\dot{\phi}^{2}+V(\phi)+\rho_{\chi}+\rho_{b}+\rho_{r} \ ,\\
\ddot{\phi}+3H\dot{\phi}+V'(\phi) &= &-\alpha'(\phi)\left(\rho_{\chi}- 3 P_{\chi}\right) \ ,\\
\ddot{\chi}+3H\dot{\chi}+e^{2\alpha(\phi)}U'(\chi) &=& -2 \alpha'(\phi)\dot{\phi}\dot{\chi} \ ,
\end{eqnarray}
where
\begin{align}
\rho_{\chi}&=e^{2\alpha(\phi)}\left[\frac{1}{2}\dot\chi^2+e^{2\alpha(\phi)}U(\chi)\right] \\
  P_{\chi}&=e^{2\alpha(\phi)}\left[\frac{1}{2}\dot\chi^2-e^{2\alpha(\phi)}U(\chi)\right]
\end{align}
are the observed dark matter energy density and pressure respectively. Comparing with~(\ref{Qcdm}) and (\ref{Qde}), we can identify 
\begin{equation}
Q=\alpha'(\phi)\dot\phi\left(\rho_{\chi}- 3 P_{\chi}\right) \ .
\end{equation}
Thus, when the field $\chi$ behaves as pressureless dark matter, $\xi=\alpha'(\phi)\dot\phi/H$. Given that the behavior at the background is that of a fluid, previously found attractors~\cite{Amendola:1999er,Bean:2008ac} will also be present here. However, it is well known that massive scalars behave as perfect fluids only at the background level, and that this behavior breaks at the level of linear perturbations due to the presence of non-adiabatic pressure. 

Since our goal is for the scalar field $\chi$ to behave as dark matter, we consider a light boson with a potential given by $U(\chi)=\frac{1}{2}m^2\chi^2$. At early times, $H\gg m$, and the field does not roll down its potential due to the large Hubble friction; in this limit, the field is frozen at its initial value. At late times, $H\ll m$, and $\chi$ behaves like a harmonic oscillator with a time dependent mass term $e^{2\alpha(\phi)}m_\chi^2$. For the purpose of examining the background solutions, we may take time averages over a time much larger than the period of oscillations of the scalar field, $T_\text{av}\gg\omega^{-1}=e^{-\alpha(\phi)}m_\chi$, and see that 
\begin{equation}
\mean{\frac{1}{2}\dot{\chi}^2}=\mean{e^{2\alpha(\phi)}U(\chi)},
\end{equation}
which shows that, on average, $\chi$ behaves as a pressureless fluid with equation of motion
\begin{equation}
\dot{{\rho}}_{\chi}+3H\rho_{\chi}=\alpha'\dot{\phi}\rho_{\chi}=Q \ . \label{eomphi}
\end{equation}
In this IDS case, considering time averages, dark matter does not redshift in the same way as standard uncoupled CDM, but instead obeys $\rho_\chi\propto e^{\alpha(\phi)}a^{-3}$. 
\begin{figure}[!ht]
	\begin{center}
		\includegraphics[scale=0.63]{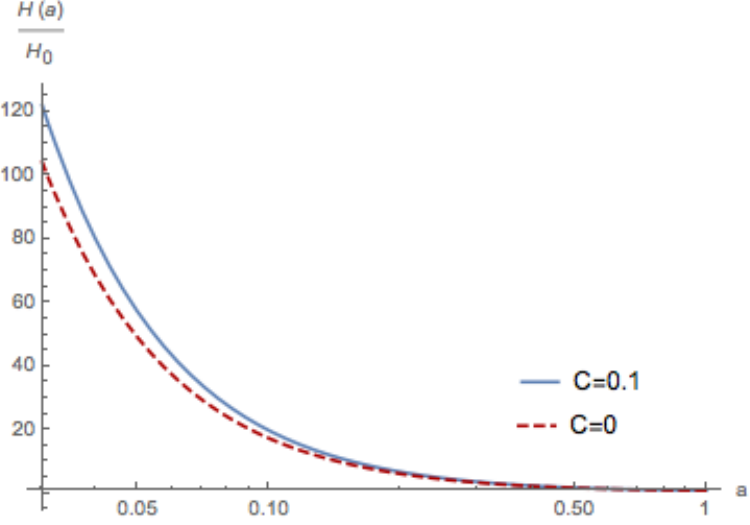}
		\includegraphics[scale=0.63]{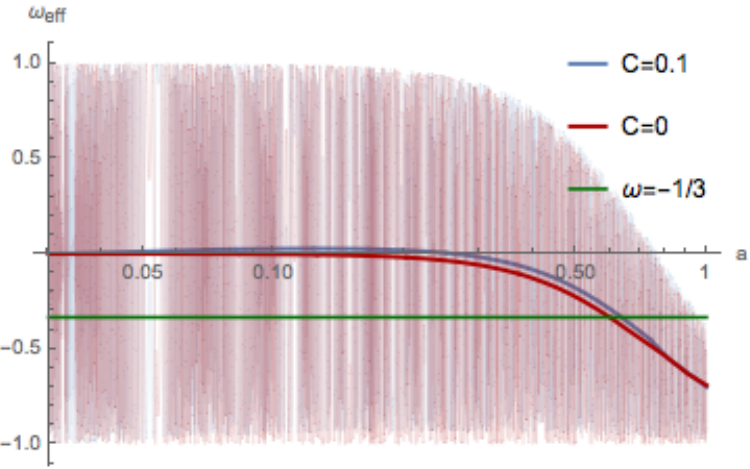} \\
		\includegraphics[scale=0.63]{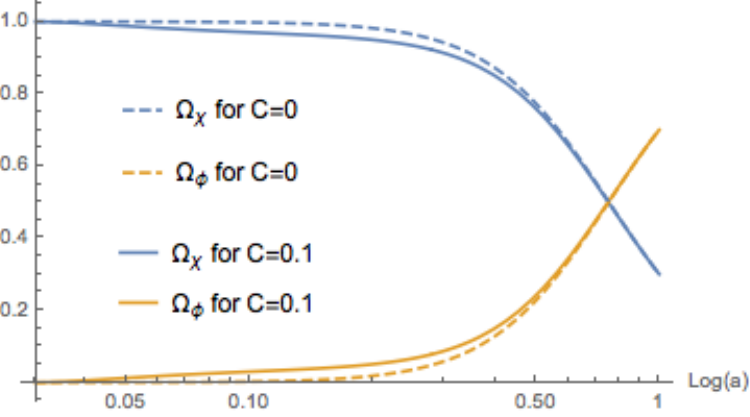}
		\caption{In this figure, we compare the coupled case (for $C=0.1$) with the uncoupled one. The top left plot compares the evolution of the Hubble parameter for a fixed $H_0$. The top right plot shows the total effective equation of state and the bottom one the behavior of the dark matter and dark energy densities.} \label{nc}
		\vspace{-0.5cm}
	\end{center}
\end{figure}

To understand the evolution completely, we solve the background equations numerically for an exponential potential $V(\phi)=V_0 e^{-\lambda\phi/M_{Pl}}$ and $\alpha=\beta \phi/M_{{Pl}}=-C \sqrt{2/3} \phi/M_{{Pl}}$, using $\lambda=0.1$ and $C=0.1$. The initial conditions and the value of $V_0$ are found by a shooting method to match the present observed cosmological parameters. The results are shown in Fig.\ref{nc}. We see that for a fixed $H_0$, $H(a)$ was larger in the past when the dark sectors interact with each other compared with the noninteracting case. From the average of the effective equation of state, we can see the transition from the matter dominated era to the accelerated one. For our fixed cosmological parameters at $a=1$, we note that in the interacting case the accelerated epoch is reached later than in the noninteracting one.

\section{Coupled $P(X)$ dark matter} \label{pofx}
\subsection{Scalar fields and non-adiabatic pressure}
In order to understand the relationship between the fluid description of IDS models and a more fundamental field theoretical description, we now seek a field theory model that reproduces the behavior of the perfect fluid models at both the background level and that of linear perturbations. In this section, we will find a rather general form for such a model. The pressure perturbations can be divided into an adiabatic and a non-adiabatic part via
\begin{equation}
\delta P=\frac{\partial P}{\partial S}\delta S+\frac{\partial P}{\partial \rho}\delta \rho=\delta P_\text{NA}+c_s^2\delta\rho \ ,
\end{equation}
where the adiabatic speed of sound is defined at zeroth order in the perturbations and can be written as $c_s^2=\dot{P}/\dot{\rho}$. In comoving gauge, the non-adiabatic pressure can be written as \cite{Unnikrishnan:2010ag}
\begin{equation}
\delta P_\text{NA}=\left(c_\phi^2-c_s^2\right)\delta\rho \ , \label{nona}
\end{equation}
where $c_\phi\equiv\tfrac{\delta P}{\delta \rho}$ is the propagation speed of the scalar fluctuations. For barotropic fluids, the non-adiabatic pressure is zero, but for scalar fields this is not always the case. As we can see from \eqref{nona}, the condition for the scalar field to behave as a barotropic fluid is $c_s^2=c_\phi^2$. In \cite{Arroja:2010wy,Unnikrishnan:2010ag,Akhoury:2008nn}, it was shown that minimally-coupled scalar fields with Lagrangians of the form\footnote{This Lagrangian can be written as a purely kinetic ``k-essence" with the field redefinition $Y=-1/2\,\partial_\mu\Phi\partial^\mu\Phi=Xg(\chi)$. If $\chi$ interacts with other fields, this redefinition may not be useful for the situation at hand.} $\lag=f(Xg(\phi))$, where $X=-1/2\,\nabla_\mu\phi\nabla^\mu\phi$, are equivalent to a barotropic fluid (assuming an irrotational fluid flow) and thus have a vanishing non-adiabatic pressure to all orders in perturbation theory. In general, it is important to take into account that the existence of interactions with other fields could lead to violations of the adiabaticity condition. However, here we will consider a conformal coupling of the form~(\ref{inter}), which leads to an insignificant violation of the adiabaticity condition, as we will see.

We begin by considering a field $\chi$ with Lagrangian of the form $\lag_\chi=f(X\,h(\chi))$ and later, we will assume that the background $\chi$ behaves as pressureless dark matter. If this field interacts with the dark energy $\phi$ as in~\eqref{inter}, then the background pressure and density are given by
\begin{align}
P_\chi&=e^{4\alpha(\phi)}f(X\,h(\chi)),\\
\rho_\chi&=e^{4\alpha(\phi)}\left[2X\frac{\partial f(X\,h(\chi))}{\partial X}-f(X\,h(\chi))\right] \ ,
\end{align}
where $X$ is now defined as $X=-\frac{1}{2}e^{-2\alpha(\phi)}\nabla_\mu\chi\nabla^\mu\chi$ to take account of the nonminimal coupling between $\chi$ and $\phi$. With these definitions, the $\chi$ field equation of motion reads
\begin{equation}
\dot{{\rho}}_\chi+3H\left(\rho_\chi+ P_\chi\right)=\alpha'\dot{\phi}\left({\rho}_\chi-3{{P}}_\chi\right)\ . 
\end{equation}
We now analyze the degree to which adiabaticity is violated due to interactions. We compute the ratio of non-adiabatic pressure $\delta P_\text{NA}$ to the total pressure perturbation $\delta P$ in comoving gauge and find
\begin{eqnarray}
\frac{\delta P_\text{NA}}{\delta P} &=& 1-\frac{c_s^2}{c_\chi^2} \nonumber \\
&=& 2\frac{\frac{\partial \lag}{\partial \chi}-2X\frac{\partial^2 \lag}{\partial X \partial \chi}+\frac{\partial \lag}{\partial \chi}\frac{2X\frac{\partial^2 \lag}{\partial X \partial X}}{\frac{\partial \lag}{\partial X}}}{ \dot{\chi}\frac{\partial \lag}{\partial X}\left(-3 H+2\dot{\alpha}\left[2\frac{\omega_\chi}{(\omega_\chi +1)}-1\right]\right)}+\dot{\alpha}\frac{4\frac{\omega_\chi}{(\omega_\chi +1)}\left(\frac{1}{c_\chi^2}-3\right)-2}{-3H+2\dot{\alpha}\left[2\frac{\omega_\chi}{(\omega_\chi +1)}-1\right]} \ . \label{pnona}
\end{eqnarray} 
Here we have used the Einstein equations and the scalar field equation of motion. We have also defined $\omega_\chi=P_\chi \big/ \rho_\chi$ and used that 
\begin{equation}
c_\chi^2=\frac{\frac{\partial \lag}{\partial X}}{\frac{\partial \lag}{\partial X}+2X\frac{\partial^2 \lag}{\partial X^2}} \ .
\end{equation}
The first term in~\eqref{pnona} vanishes for a Lagrangian of the form we are considering $\lag_\chi=g(\phi)f(X\,h(\chi))$ by a simple extension of the arguments in~\cite{Arroja:2010wy,Unnikrishnan:2010ag}. In the following, we discuss the case in which the field behaves as dark matter and thus satisfies
\begin{equation}
\frac{P_\chi}{\rho_\chi}\rightarrow0,\qquad\frac{\rho_\chi+P_\chi}{2X\frac{\partial\rho_\chi
	}{\partial X}}\rightarrow0 \ ,
\end{equation}
where the second equation is simply the requirment that $c_\chi^2\rightarrow0$. In this case, the equation of motion for the field $\chi$ is approximately that of a pressureless fluid
\begin{equation}
\dot{{\rho}}_\chi+3H\rho_\chi=\alpha'\dot{\phi}{\rho}_\chi \ . \label{eomx}
\end{equation}
For this pressureless field, we then find that
\begin{equation}
\frac{\delta P_\text{NA}}{\delta P}\sim\frac{1-\left(\frac{4\omega_\chi}{3c_\chi^2}\right)}{\frac{H}{\dot{\alpha}}+\frac{2}{3}} .
\end{equation}
Furthermore, if we consider a coupling of the form $\alpha(\phi)=\beta\phi \big/ M_{Pl}$, we have $|\dot\alpha|\lesssim\beta \sqrt{\rho_\phi} \big /M_{Pl}$. Using this, and the fact that $H=\sqrt{\rho_{\chi}+\rho_\phi} \big /M_{Pl}$, yields the requirement
\begin{equation}
\left|\frac{\delta P_\text{NA}}{\delta P}\right|\lesssim \beta\left|1-\frac{4}{3}\frac{\omega_\chi}{c_\chi^2}\right| \ .
\end{equation}
For known $P(X)$ Lagrangians that lead to a pressureless field, such as DBI or $X^n$ with large $n$, we have that $|\omega_\chi|=|c_\chi^2|$. In such cases, this means that the non-adiabatic pressure is negligible as long as the coupling is small. Thus, we have shown that in the cases of interest $\delta P_\text{NA}\big/ \delta P\ll1$.

\subsection{DBI dark matter coupled to dark energy}
One of the simplest Lagrangians that satisfies the conditions to behave as fluid dark matter is $\lag= M^4 X^n$ for large $n$. In this case we have $w=c_\chi^2=1/(2n-1)$.
Another more sophisticated option, which we will analyze in detail, is the DBI Lagrangian. This Lagrangian describes the motion of a 3-brane in a 5 dimensional spacetime, and reads
\begin{equation}
\lag=-M^4\sqrt{1-2X} \ ,
\end{equation}
where $M$ is a mass scale corresponding to the brane tension. We will consider a conformal coupling, leading to the dark matter action 
\begin{equation}
S_\chi=-\int\ud ^4 x \sqrt{-g}\; e^{4\alpha(\phi)} M^4\sqrt{1-2e^{-2\alpha(\phi)}X} \ ,
\end{equation}
where $X=\frac{1}{2}\nabla_\mu\chi\nabla^\mu\chi$ is now defined with respect to the metric whose geodesics are followed by the Standard Model matter. Defining
\begin{equation}
\gamma\equiv \frac{1}{\sqrt{1-2e^{-2\alpha(\phi)}X}} \ , \label{gam}
\end{equation} 
we may write the equation of state and speed of fluctuations respectively as
\begin{equation}
\omega=-\frac{1}{\gamma^2},\qquad c_\chi^2=\frac{1}{\gamma^2} \ .
\end{equation}
This shows that, as long as we are in the ``relativistic limit", in which $\gamma\gg1$, the field $\chi$ behaves approximately as pressureless dark matter. It is interesting to notice that the field evolves towards $\gamma\sim 1$, so $\chi$ will start behaving as dark energy in the future. The equation of motion for $\chi$ is given by
\begin{equation}
\frac{\ud}{\ud t}\left(a^3\frac{e^{2\alpha(\phi)}\sqrt{2X}}{\sqrt{1-2e^{-2\alpha(\phi)}X}}\right)=0 \ ,
\end{equation}
which allows us to express $\gamma$ on-shell as
\begin{equation}
\gamma=\sqrt{1+\frac{A^2}{a^6}e^{-6\alpha(\phi)}} \ , \label{gamma}
\end{equation}
where $A$ is a constant that we may fix by demanding that the energy density in the $\chi$ field matches the observed abundance of dark matter today. In order for $\chi$ to behave as approximately pressureless dark matter today, we must require $(A^2/a^6) e^{-6\alpha(\phi)} \gg 1$. However, based on previous studies we expect that broad consistency with cosmological observations will require $\alpha_0 \equiv \alpha(\phi(t_0)) \lesssim 1$. Thus, $A$ is the quantity primarily responsible for determining whether $\chi$ behaves as CDM. Note also that, for this model, the dark matter density redshifts as 
\begin{equation}
\rho=M^4e^{4\alpha(\phi)}\gamma\sim M^4e^{\alpha(\phi)}\frac{A}{a^3}\ . \label{rhochi}
\end{equation}
Using this, and requiring $\rho(\text{today})=\rho_\text{CDM}^\text{observed}$, we find that the constant $A$ is given by
\begin{equation}
A=3\Omega_\text{CDM}\,e^{-\alpha(\phi)}\frac{H_0^2 M_\text{Pl}^2}{M^4} \sim 30 \left(\frac{2.7\times 10^{-5}\text{ eV}}{M}\right)^4 \ . \label{const}
\end{equation}
For $A=30$, we obtain $-\omega_\chi=c_\chi^2\sim10^{-3}$. Thus, a brane tension $M\sim10^{-5}$ eV or smaller would give the desired behavior. The expression~\eqref{const} for $A$ is a general result for any $P(X)$ theory that behaves as fluid dark matter. Although in the DBI case this results in a limit on the scale $M$ in order that the theory behaves as dark matter, in other cases, for example in the $X^n$ case, there is no restriction on $M$. Given that not only the background, but also the linear perturbations behave as the fluid case; this means that, for $\alpha=\beta \phi/M_{Pl}$ with  $\beta\sim0.066$, the $H_0$ tension is alleviated in these models, by construction.

\subsection{Quantum corrections}
\begin{figure}[!ht]
	\begin{center}
		\includegraphics[scale=0.8]{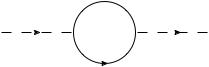} \includegraphics[scale=0.8]{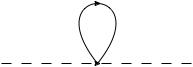} 
		\caption{Loop corrections to the dark energy mass from a dark matter loop. Both graphs give the same contribution.} \label{mass}
		\vspace{-0.5cm}
	\end{center}
\end{figure}

In order to analyze the quantum corrections, we will use power counting techniques to estimate the amplitudes following \cite{Burgess:2007pt,Goon:2016ihr}.
We begin by examining the requirements to keep quantum corrections to the dark energy mass under control. Consider the quantum corrections to the dark energy field amplitudes given by $\chi$ loops. The loop corrections to the n-point scattering amplitude for $\phi$ are given by
\begin{equation}
\mathcal{M}^{(n)}_{\phi}=\beta^n\frac{M^4}{M_{Pl}^n}\left(\frac{k}{M}\right)^{2L+2+\sum_m(m-2)V_m} \ ,
\end{equation}
where $L$ is the number of $\chi$ loops, $V_m$ is the number of vertices with m $\chi$ lines, and $M$ is the strong coupling scale of the theory. From this we find that the largest quantum correction to the dark energy mass is given by
\begin{equation}
\Delta m_\phi=\beta\frac{k^2}{M_{Pl}} \ ,
\end{equation}
which corresponds to two $\phi$'s interacting through one $\chi$ loop, as seen in Fig.\ref{mass}. Requiring that this correction is small gives
\begin{equation}
k<\left(\frac{m_\phi M_{Pl}}{\beta}\right)^{1/2}\sim\left(\frac{10^{-2}}{\beta}\right)0.1\  \text{eV}
\end{equation}
However, in order to be self-consistent, any calculation we do must be below the cutoff of the theory;  therefore, we will always work in the regime $k\leq \Lambda_\text{c}$, where $\Lambda_\text{c}$ is the cutoff. In principle, the cutoff can be equal to or larger than the strong coupling scale $\sim M\sim10^{-5}$ eV. As long as the cutoff is such that $\Lambda_\text{c}\leq0.1$ eV and $\beta$ is not too large (which is forbidden by other constraints), the corrections will be under control.

More generally, we can consider the amplitudes involving $i$ copies of the $\chi$ field and $j$ copies of the $\phi$ field. To do so, we write the dark sector Lagrangian as an infinite power series 
\begin{equation}
\lag=-\frac{1}{2}\left(\nabla\phi\right)^2-V(\phi) +\sum_{n,l,m=0}^{\infty} C_{n,l,m} M^4 \left(-\frac{(\nabla\chi)^2}{M^4}\right)^{m}\left(\frac{\beta \phi}{M_{Pl}}\right)^{n+l} \ , \label{series}
\end{equation}
where
\begin{equation}C_{n,l,m}=\binom{\frac{1}{2}}{m}\frac{1}{n!\,l!}4^n(-2m)^l \ ,
\end{equation} 
and we have reintroduced the corresponding mass scales to ensure a canonically normalized kinetic term and fields with mass dimension one. In the following, it is interesting to note that the results do not depend strongly on the coefficients unless fine-tuning exists. For simplicity, we will analyze the concrete example of an exponential potential for the dark energy field, but the calculation for any other potential is straightforward. We expand the self-interacting potential of $\phi$ as
\begin{equation}
V(\phi)=V_0e^{-\lambda\phi/M_{Pl}}=V_0\sum_{k=0}^{\infty}\frac{(-1)^k}{k!}\left(\frac{\lambda\, \phi}{M_{Pl}}\right)^{k} \ .
\end{equation}
Given the above considerations, the scattering amplitude reads
\begin{equation}
\mathcal{M}\simeq\left(\frac{1}{M}\right)^{a-4}\left(\frac{\beta}{M_{Pl}}\right)^{b}\left(\frac{q}{M}\right)^{c}\left(\frac{\lambda \; V_0}{\beta\; M^4}\right)^{\sum n S_n} \ , \label{ampl}
\end{equation}
with
\begin{align}
&a=\sum i \,V_{i\,j}-2 I, \qquad b=\sum j\, V_{i\,j} \nonumber \\
 &c=2L+2+\sum (i-2) \,V_{i\,j} \ ,\label{amp}
\end{align}
where $V_{i\,j}$ is the number of vertices with $i$ $\chi$ and $j$ $\phi$ lines attached to it, $I=I_\phi+I_\chi$ is the number of internal lines, $L$ is the number of loops, and $S_n$ is the number of self-interaction $\phi$ vertices with $n$ $\phi$ fields. The momentum $q$ stands for a combination of the external momenta; this is the dominant scale given that we are working with light fields. In the following, we will point out in which situations we should replace some of the $q$ factors appearing in Eq. \eqref{ampl} by a different mass scale. In all of these cases, the external momenta appearing in the actual expression for a factor $q$ (computed by using Feynman rules and without any approximation) are small and thus subdominant. In such situations, we should replace $q$ by the dominant scale that enters in the corresponding momentum factor. This will reduce the powers of $q$ appearing in Eq.\eqref{ampl}.
	
If one of the external momenta is less than or of order $m_\phi\sim H$, then $m_\phi$ will become the dominant scale. In the momentum factors where this small momentum is involved, we should replace the corresponding $q$ by $m_\phi$ in Eq.\eqref{ampl}. The results stated above were computed in flat space. A more precise calculation would require the use of the scalar field propagator in an FRW space. While the flat space analysis is accurate when all the external momenta are larger than $H$, if some of the external momenta are smaller than $H$, then some $q$'s will be replaced by $H$. Note also that, if all momenta are smaller than $H$ then we can integrate out the field $\phi$ entirely (since $m_\phi\sim H_0$). The leading order contributions from integrating out $\phi$ are of order $M^4\frac{M^4}{M_{Pl}^2\,m_\phi^2}\sim 10^{-28} \text{eV}^{4} \left(M / 10^{-5} \text{eV}\right)^8$, demonstrating that, as expected, in this regime the theory is effectively an uncoupled DBI dark matter model. 

It is well known that for a DBI field classical solutions with $X\sim1$ are valid as long as the acceleration is small: $\ddot{\chi}\big/M^3\ll\gamma^{-3}$ \cite{deRham:2014wfa}. In our coupled cased, it is necessary to reanalyze whether the quantum corrections to the interacting DBI sector are under control. The leading order tree-level amplitude for $n$ $\chi$ fields and $m$ $\phi$ fields is
\begin{equation}
\mathcal{M}_\text{tree}^{(n+m)}=M^{4-n}\left(\frac{\beta}{M_{Pl}}\right)^m\left(\frac{q}{M}\right)^n \ .
\end{equation}
The other contribution to the tree level amplitude comes from two vertices interacting through a $\phi$ field. This amplitude is suppressed by $\beta\left(M/(q M_{Pl})\right)^2$.
From~\eqref{amp}, we see that some loop amplitudes involving $\phi$ can renormalize the $P(X)$ terms. Nevertheless, these contributions are under control; every $\phi$ propagator in a loop gives a contribution suppressed by $\beta\left(M/M_{Pl}\right)^2\sim10^{-66}$. All other contributions contain higher powers of momentum and do not renormalize the original interactions.  If some of the external momenta are smaller than $m_\phi\sim H_0$ or $H$, some of the $q$'s will be replaced by $m_\phi$ or $H$. These contributions will be suppressed by $\frac{m_\phi}{M}\sim\frac{H}{m}\sim10^{-28}$.  We therefore conclude that the quantum corrections are under control and we may trust the classical solution with large velocity ($X\sim1$) as long as the acceleration is small.

As we mentioned earlier, this analysis does not depend on the coefficients of the series expansion~\eqref{series} (unless these are fine-tuned). This means that the quantum corrections are under control for a generic $P(X)$ theory coupled to dark energy as in~\eqref{inter}, as long as higher derivative terms are small and $M\ll M_{Pl}$.

\subsection{Validity of the linear regime: fields vs fluids}
Thus far, we have worked at the level of the background cosmology and its linear perturbations, and have examined what is required for the fluid and field theoretical formulations to yield the same results in that regime. We now consider the differences between these descriptions that may arise at nonlinear scales, when the field gradients become large and, for example, non-canonical scalar fields can develop caustics~\cite{Felder:2002sv,deRham:2016ged,Mukohyama:2016ipl,Babichev:2016hys}. This is important because if caustics form, then a UV completion of the theory is needed in order to obtain any conclusions in this regime. It has been proposed that a $P(X)$ field admits a caustic-free completion by means of a canonical complex scalar field \cite{Babichev:2017lrx}, although the validity in the interacting case has not been established. For some models of DBI-like dark matter, the linear theory is valid only for a small period of time. Specifically, models with a Lagrangian $\lag=-V(\chi)\sqrt{1-2X}$, with $V(\chi)=V_0 e^{\chi/\chi_0}$, behave as pressureless matter and can be considered a dark matter candidate, but the linear regime breaks down when $t>10 \chi_0^{-1}$ \cite{Frolov:2002rr}. Here we analyze if and when such an effect takes place in our model.

We compute the perturbations in Newtonian gauge for a weakly coupled $P(X)$ theory that behaves as dark matter. The perturbed metric is given by
\begin{equation}
\ud s^2=-\left(1+2\Phi\right)\ud t^2+\left(1-2\Phi\right)a^2(t)\ud \mathbf{x}^2 \ .
\end{equation} 
We work in the dark matter dominated epoch, where the $\phi$ energy density is sub-dominant and thus can be neglected. The two independent perturbed Einstein equations are
\begin{align}
\frac{(a \Phi)^{.}}{a}&=\frac{1}{2 M_{Pl}^2}(\rho_\chi+P_\chi)\frac{\xi}{\dot{\chi}} \ ,\\
\left(\frac{\xi}{\dot\chi}\right)^{.}-\frac{2\alpha_\phi\varphi}{\dot{\chi}^2}&=\left[1+\frac{2 c_\chi^2  M_{Pl}^2}{a^2 (\rho_\chi+P_\chi)}\nabla^2\right]\Phi \ ,
\end{align}
where $\xi$ is the DM field perturbation, an overdot denotes a derivative with respect to $t$, and the subscript $\phi$ a derivative with respect to $\phi$. We now introduce the Sasaki-Mukhanov variable $\nu$ defined through
\begin{equation}
\frac{\nu}{z}=\frac{5\rho_\chi+3P_\chi}{3(\rho_\chi+P_\chi)}\Phi+\frac{2\rho_\chi}{3(\rho_\chi+P_\chi)}\frac{\dot\Phi}{H} \ ,
\end{equation}
where
\begin{equation}
z=\frac{a \sqrt{\rho_\chi+P_\chi}}{c_\chi H} \ .
\end{equation}
Note that we have assumed that $e^{\alpha(\phi)}$ is constant, which is justified since this quantity varies slowly during matter domination. By solving the equations numerically, we have checked that this approximation will only introduce an error of order $\lesssim10$\%. Given these definitions, we may write the two perturbed Einstein equations as one second order equation for $\nu$
\begin{equation}
\nu''+\left(c_\chi^2\nabla^2\left(1+\frac{2}{\nu}\int\frac{\alpha_\phi \mathcal{H}}{X \, e^{\alpha}}\,\varphi\,\ud \eta \right) -\frac{z''}{z}\right)\nu=0 \ , \label{nueq}
\end{equation}
where a prime now denotes a derivative with respect to conformal time $\eta$, defined via $dt^2=a(\eta)^2d\eta^2$ and $\mathcal{H}=\dot{a}$ is the conformal Hubble parameter. We will solve this equation in two limits: the short wavelength limit and the long wavelength limit. We start by analyzing the long wavelength limit defined by $k c_\chi\ll a H$. In this limit, the term proportional to the speed of sound is negligible  which means that $\nu=z$ is a solution. Using this and the fact that $H\propto t^{-1}$, we find that $\Phi(\mathbf{x},t)=\Phi(\mathbf{x})+\Phi_\text{d}(\mathbf{x}) \,t^{-5/3}$, which resembles the CDM case. This is not surprising because we know our field describes an almost pressureless fluid in this limit. In order to find the behavior of the DM field perturbation $\xi$ we look at the $(0\,j)$ component of the Einstein equations and take $a\propto e^{\alpha(\phi)/3} \,t^{2/3}$. Neglecting the decaying mode $\Phi(\mathbf{x})_\text{d}$, we find that 
\begin{equation}
\xi(\mathbf{x},t)=\dot{\chi}\Phi(\mathbf{x})t \ .
\end{equation}
As long as
\begin{align}
\Phi(\mathbf{x})\left(1+\frac{\ddot{\chi}t}{\dot{\chi}}\right)+\nabla\Phi\,t<1,
\end{align}
is satisfied, the linear regime for the field perturbations is valid. For a small acceleration and large velocity, as needed for quantum corrections to be under control, the second term is small. The third term can become large at small distances; however, at these distances the long wavelength solution may no longer be valid. If this is the case, we should look at the short scales solution to find when the linear regime breaks.

\begin{figure}[!ht]
	\begin{center}
		\includegraphics[scale=0.77]{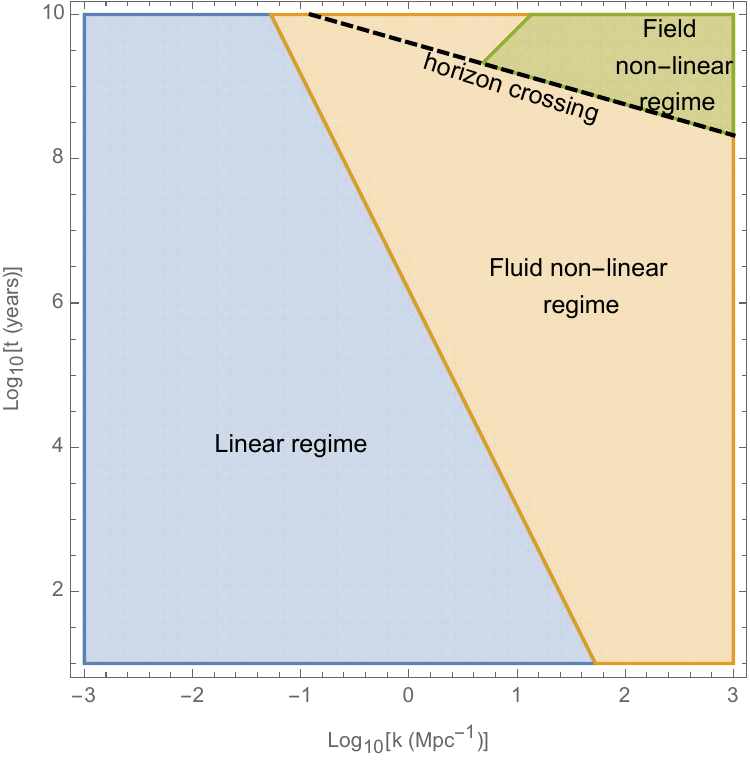} 
		\caption{Approximate regions of linear and nonlinear regime in $k-t$ plane for the coupled DBI model with $A=30$. The dotted line shows the horizon crossing: $k c_\chi=a H$. The fluid nonlinear regime starts when $3/2 \, k^2\Phi a_0^{-2}t_0^{4/3} t^{2/3}>1$. On the other hand, the field nonlinear regime starts when the physical wavelength of its perturbation is smaller than the sound horizon and $k\Phi\,t_0^{5/3}t^{-2/3}>1$. } \label{nonlin}
	\end{center}
\vspace{-0.5cm}
\end{figure}

For short wavelengths, we have $k c_\chi\gg a H$, which means that we can neglect the term $\frac{z''}{z}\sim a^2 H^2$. We also neglect the term corresponding to the dark energy perturbation since this is negligible during matter domination.  In order to analyze this limit, we will focus on the coupled DBI case. For this case, we have that $c_\chi\sim a^3/A$ and thus
\begin{eqnarray}
x\equiv\frac{k c_\chi}{a H}\sim\frac{k \,a_0^2\, t^{7/3}}{t_0^{4/3}A},
\end{eqnarray}
where $
a_0$ and $t_0$ are the scale factor and time at which matter domination ends respectively. We find that at short scales the solution to~\eqref{nueq} is $\nu\sim x^{-3/7}\cos\left(3\pi / 4-x \right)$ which in turn leads to $\Phi(\mathbf{x},t)=\Phi_\text{d}(\mathbf{x}) \,\left(t \big/ t_0\right)^{-5/3}+\Phi_\text{sub-leading}(\mathbf{x,t})$. Given this solution, we find that the linear regime for the field perturbations is valid as long as 
\begin{align}
\Phi_d(\mathbf{x})\left(\frac{\ddot{\chi}t_0}{\dot{\chi}}\left(\frac{t_0}{t}\right)^{2/3}-\frac{2}{3}\left(\frac{t_0}{t}\right)^{5/3}\right)+\nabla\Phi_d\,t_0\left(\frac{t_0}{t}\right)^{2/3}<1.
\end{align}
This expression is only valid at short scales where the first two terms are small. Only the last term will grow large at small scales and break the validity of the linear regime, see Fig~\ref{nonlin}.

It is instructive to also analyze the situation from the fluid point of view and identify the circumstances under which the energy density perturbation becomes large. Using the $(0,0)$ and $(0,i)$ perturbed Einstein equations, we can write the fluid perturbation as
\begin{equation}
\delta=\frac{2 M_{Pl}^2}{\rho}\left[\frac{1}{a^2}\nabla^2\Phi-\frac{3H}{2 M_{Pl}^2}(\rho_\chi+P_\chi)\frac{\xi}{\dot{\chi}}\right] \ .
\end{equation}
Notice that only the second term depends on the matter content of the Universe. The overdensity can be rewritten as 
\begin{equation}
\delta=\tfrac{3}{2 a_0^2}t_0^{4/3} \nabla^2\Phi t^{2/3}-2\Phi+\text{decaying modes} \ .
\end{equation}
This demonstrates that the validity of the fluid linear regime is the same as in the CDM case, as expected. Since the gradients are suppressed at distances larger than the Hubble radius the density perturbation is constant in that regime. At smaller distances, the gradients become larger and the density perturbation grows.

Interestingly, we can see that the fluid nonlinear regime is reached before the field gradients grow large, as can be seen in Fig.\ref{nonlin}. This means that, the field theory can be trusted to perform calculations in the fluid nonlinear regime without worrying about the formation of caustics. In the DBI case, for a given scale $k$ satisfying $k\Phi\,t_0^{5/3}t^{-2/3}>1$ inside the sound horizon, caustics could form and we cannot trust any conclusions drawn in the nonlinear limit. If the DBI field $\chi$ represents the position of a brane in a higher dimension, these multivalued regions correspond to the folding of the brane and can be removed by choosing a different slicing of the extra dimension. However, in the interacting case, such a slicing may not exist. Previous results have shown that for a non-interacting DBI field, both planar and spherical waves in Minkowski space~\cite{Mukohyama:2016ipl,deRham:2016ged} evolve in a caustic-free way; it is unknown if this result holds in the interacting case where gravity is tuned on.

\section{Conclusions}\label{conclu}
Interactions between the different components of the dark sector can naturally arise from higher dimensional theories or Brans-Dicke-like theories. A range of phenomenological models for an interacting dark sector have been constructed and analyzed; these models usually treat dark matter (and even dark energy) as a barotropic fluid and have most recently been suggested as a way to address the tension between the $H_0$ values inferred from observations at different redshifts. Here, we have focused on a more fundamental perspective by constructing field theory models which behave as the fluid ones even in a large region of the nonlinear regime.

Since a thermal fermion WIMP will lead to large quantum corrections to the dark energy mass, we have focused on scalar dark matter. We have extended previous work to propose a field theory model which behaves as a pressureless fluid at the background level, and also up to linear order in the perturbations. This model consists of a $P(X)$ field, $\chi$, conformally coupled to the dark energy. Focusing specifically on the case of a DBI field, we have imposed an upper limit on the DBI tension in order to assure that $\chi$ behaves as dark matter up to the present epoch. We have shown that quantum corrections to these kinds of interacting models are under control for a large velocity as long as the associated higher derivative terms are small. Finally, we have analyzed the validity of the linear regime in these models. While the validity of the fluid linear regime is the same as in the uncoupled CDM case, we have shown that the linear regime for the field theory breaks down much later. This implies that we need not worry about the formation of caustics in a larger region of the fluid nonlinear regime. When $\chi$ gradients grow large, caustics could form, but this behavior takes place out of the regime of validity of the effective field theory.\\

\paragraph{Acknowledgments:}
We thank Sina Bahrami, Benjamin Elder, Eanna Flannagan, Justin Khoury, Vinicius Miranda, Lucas Secco and Adam Solomon for helpful discussions. We also thank Joseph P J for pointing out issues with the dark matter equations in Sec. 3 on a previous version of this paper. The work of M.C. and M.T was supported in part by NASA ATP grant NNX11AI95G. M.T. was also supported in part by US Department of Energy (HEP) Award DE-SC0013528.

\renewcommand{\em}{}
\bibliographystyle{utphys}
\bibliography{bibliography}

\providecommand{\href}[2]{#2}\begingroup\raggedright\begin{thebibliography}{10}

\bibitem{Riess:2016jrr}
A.~G. Riess {\em et al.}, ``{A 2.4\% Determination of the Local Value of the
  Hubble Constant},'' \href{http://dx.doi.org/10.3847/0004-637X/826/1/56}{{\em
  Astrophys. J.} {\bf 826} (2016) no.~1, 56},
\href{http://arxiv.org/abs/1604.01424}{{\tt arXiv:1604.01424 [astro-ph.CO]}}.

\bibitem{Ade:2015xua}
{\bf Planck} Collaboration, P.~A.~R. Ade {\em et al.}, ``{Planck 2015 results.
  XIII. Cosmological parameters},''
\href{http://arxiv.org/abs/1502.01589}{{\tt arXiv:1502.01589 [astro-ph.CO]}}.

\bibitem{DiValentino:2017iww}
E.~Di~Valentino, A.~Melchiorri, and O.~Mena, ``{Can interacting dark energy
  solve the $H_0$ tension?},''
  \href{http://dx.doi.org/10.1103/PhysRevD.96.043503}{{\em Phys. Rev.} {\bf
  D96} (2017) no.~4, 043503},
\href{http://arxiv.org/abs/1704.08342}{{\tt arXiv:1704.08342 [astro-ph.CO]}}.

\bibitem{Salvatelli:2013wra}
V.~Salvatelli, A.~Marchini, L.~Lopez-Honorez, and O.~Mena, ``{New constraints
  on Coupled Dark Energy from the Planck satellite experiment},''
  \href{http://dx.doi.org/10.1103/PhysRevD.88.023531}{{\em Phys. Rev.} {\bf
  D88} (2013) no.~2, 023531},
\href{http://arxiv.org/abs/1304.7119}{{\tt arXiv:1304.7119 [astro-ph.CO]}}.

\bibitem{Pettorino:2013oxa}
V.~Pettorino, ``{Testing modified gravity with Planck: the case of coupled dark
  energy},'' \href{http://dx.doi.org/10.1103/PhysRevD.88.063519}{{\em Phys.
  Rev.} {\bf D88} (2013)  063519},
\href{http://arxiv.org/abs/1305.7457}{{\tt arXiv:1305.7457 [astro-ph.CO]}}.

\bibitem{Pettorino:2012ts}
V.~Pettorino, L.~Amendola, C.~Baccigalupi, and C.~Quercellini, ``{Constraints
  on coupled dark energy using CMB data from WMAP and SPT},''
  \href{http://dx.doi.org/10.1103/PhysRevD.86.103507}{{\em Phys. Rev.} {\bf
  D86} (2012)  103507},
\href{http://arxiv.org/abs/1207.3293}{{\tt arXiv:1207.3293 [astro-ph.CO]}}.

\bibitem{Amendola:1999er}
L.~Amendola, ``{Coupled quintessence},''
  \href{http://dx.doi.org/10.1103/PhysRevD.62.043511}{{\em Phys. Rev.} {\bf
  D62} (2000)  043511},
\href{http://arxiv.org/abs/astro-ph/9908023}{{\tt arXiv:astro-ph/9908023
  [astro-ph]}}.

\bibitem{Bean:2008ac}
R.~Bean, E.~E. Flanagan, I.~Laszlo, and M.~Trodden, ``{Constraining
  Interactions in Cosmology's Dark Sector},''
  \href{http://dx.doi.org/10.1103/PhysRevD.78.123514}{{\em Phys. Rev.} {\bf
  D78} (2008)  123514},
\href{http://arxiv.org/abs/0808.1105}{{\tt arXiv:0808.1105 [astro-ph]}}.

\bibitem{Koivisto:2005nr}
T.~Koivisto, ``{Growth of perturbations in dark matter coupled with
  quintessence},'' \href{http://dx.doi.org/10.1103/PhysRevD.72.043516}{{\em
  Phys. Rev.} {\bf D72} (2005)  043516},
\href{http://arxiv.org/abs/astro-ph/0504571}{{\tt arXiv:astro-ph/0504571
  [astro-ph]}}.

\bibitem{Amendola:2003eq}
L.~Amendola and C.~Quercellini, ``{Tracking and coupled dark energy as seen by
  WMAP},'' \href{http://dx.doi.org/10.1103/PhysRevD.68.023514}{{\em Phys. Rev.}
  {\bf D68} (2003)  023514},
\href{http://arxiv.org/abs/astro-ph/0303228}{{\tt arXiv:astro-ph/0303228
  [astro-ph]}}.

\bibitem{Amendola:2003wa}
L.~Amendola, ``{Linear and non-linear perturbations in dark energy models},''
  \href{http://dx.doi.org/10.1103/PhysRevD.69.103524}{{\em Phys. Rev.} {\bf
  D69} (2004)  103524},
\href{http://arxiv.org/abs/astro-ph/0311175}{{\tt arXiv:astro-ph/0311175
  [astro-ph]}}.

\bibitem{Amendola:2002bs}
L.~Amendola, C.~Quercellini, D.~Tocchini-Valentini, and A.~Pasqui,
  ``{Constraints on the interaction and selfinteraction of dark energy from
  cosmic microwave background},'' \href{http://dx.doi.org/10.1086/368064}{{\em
  Astrophys. J.} {\bf 583} (2003)  L53},
\href{http://arxiv.org/abs/astro-ph/0205097}{{\tt arXiv:astro-ph/0205097
  [astro-ph]}}.

\bibitem{Bean:2007nx}
R.~Bean, E.~E. Flanagan, and M.~Trodden, ``{The Adiabatic Instability on
  Cosmology's Dark Side},''
  \href{http://dx.doi.org/10.1088/1367-2630/10/3/033006}{{\em New J. Phys.}
  {\bf 10} (2008)  033006},
\href{http://arxiv.org/abs/0709.1124}{{\tt arXiv:0709.1124 [astro-ph]}}.

\bibitem{Bean:2007ny}
R.~Bean, E.~E. Flanagan, and M.~Trodden, ``{Adiabatic instability in coupled
  dark energy-dark matter models},''
  \href{http://dx.doi.org/10.1103/PhysRevD.78.023009}{{\em Phys. Rev.} {\bf
  D78} (2008)  023009},
\href{http://arxiv.org/abs/0709.1128}{{\tt arXiv:0709.1128 [astro-ph]}}.

\bibitem{LaVacca:2009yp}
G.~La~Vacca, J.~R. Kristiansen, L.~P.~L. Colombo, R.~Mainini, and S.~A.
  Bonometto, ``{Do WMAP data favor neutrino mass and a coupling between Cold
  Dark Matter and Dark Energy?},''
  \href{http://dx.doi.org/10.1088/1475-7516/2009/04/007}{{\em JCAP} {\bf 0904}
  (2009)  007},
\href{http://arxiv.org/abs/0902.2711}{{\tt arXiv:0902.2711 [astro-ph.CO]}}.

\bibitem{Pettorino:2008ez}
V.~Pettorino and C.~Baccigalupi, ``{Coupled and Extended Quintessence:
  theoretical differences and structure formation},''
  \href{http://dx.doi.org/10.1103/PhysRevD.77.103003}{{\em Phys. Rev.} {\bf
  D77} (2008)  103003},
\href{http://arxiv.org/abs/0802.1086}{{\tt arXiv:0802.1086 [astro-ph]}}.

\bibitem{Wintergerst:2010ui}
N.~Wintergerst and V.~Pettorino, ``{Clarifying spherical collapse in coupled
  dark energy cosmologies},''
  \href{http://dx.doi.org/10.1103/PhysRevD.82.103516}{{\em Phys. Rev.} {\bf
  D82} (2010)  103516},
\href{http://arxiv.org/abs/1005.1278}{{\tt arXiv:1005.1278 [astro-ph.CO]}}.

\bibitem{Mainini:2006zj}
R.~Mainini and S.~Bonometto, ``{Mass functions in coupled Dark Energy
  models},'' \href{http://dx.doi.org/10.1103/PhysRevD.74.043504}{{\em Phys.
  Rev.} {\bf D74} (2006)  043504},
\href{http://arxiv.org/abs/astro-ph/0605621}{{\tt arXiv:astro-ph/0605621
  [astro-ph]}}.

\bibitem{Baldi:2010td}
M.~Baldi and V.~Pettorino, ``{High-z massive clusters as a test for dynamical
  coupled dark energy},''
  \href{http://dx.doi.org/10.1111/j.1745-3933.2010.00975.x}{{\em Mon. Not. Roy.
  Astron. Soc.} {\bf 412} (2011)  L1},
\href{http://arxiv.org/abs/1006.3761}{{\tt arXiv:1006.3761 [astro-ph.CO]}}.

\bibitem{Baldi:2010vv}
M.~Baldi, ``{Time dependent couplings in the dark sector: from background
  evolution to nonlinear structure formation},''
  \href{http://dx.doi.org/10.1111/j.1365-2966.2010.17758.x}{{\em Mon. Not. Roy.
  Astron. Soc.} {\bf 411} (2011)  1077},
\href{http://arxiv.org/abs/1005.2188}{{\tt arXiv:1005.2188 [astro-ph.CO]}}.

\bibitem{Saracco:2009df}
F.~Saracco, M.~Pietroni, N.~Tetradis, V.~Pettorino, and G.~Robbers,
  ``{Non-linear Matter Spectra in Coupled Quintessence},''
  \href{http://dx.doi.org/10.1103/PhysRevD.82.023528}{{\em Phys. Rev.} {\bf
  D82} (2010)  023528},
\href{http://arxiv.org/abs/0911.5396}{{\tt arXiv:0911.5396 [astro-ph.CO]}}.

\bibitem{Casas:2015qpa}
S.~Casas, L.~Amendola, M.~Baldi, V.~Pettorino, and A.~Vollmer, ``{Fitting and
  forecasting coupled dark energy in the non-linear regime},''
  \href{http://dx.doi.org/10.1088/1475-7516/2016/01/045}{{\em JCAP} {\bf 1601}
  (2016) no.~01, 045},
\href{http://arxiv.org/abs/1508.07208}{{\tt arXiv:1508.07208 [astro-ph.CO]}}.

\bibitem{Randall:1999ee}
L.~Randall and R.~Sundrum, ``{A Large mass hierarchy from a small extra
  dimension},'' \href{http://dx.doi.org/10.1103/PhysRevLett.83.3370}{{\em Phys.
  Rev. Lett.} {\bf 83} (1999)  3370--3373},
\href{http://arxiv.org/abs/hep-ph/9905221}{{\tt arXiv:hep-ph/9905221
  [hep-ph]}}.

\bibitem{Amendola:1999qq}
L.~Amendola, ``{Scaling solutions in general nonminimal coupling theories},''
  \href{http://dx.doi.org/10.1103/PhysRevD.60.043501}{{\em Phys. Rev.} {\bf
  D60} (1999)  043501},
\href{http://arxiv.org/abs/astro-ph/9904120}{{\tt arXiv:astro-ph/9904120
  [astro-ph]}}.

\bibitem{Wetterich:1994bg}
C.~Wetterich, ``{The Cosmon model for an asymptotically vanishing time
  dependent cosmological 'constant'},'' {\em Astron. Astrophys.} {\bf 301}
  (1995)  321--328,
\href{http://arxiv.org/abs/hep-th/9408025}{{\tt arXiv:hep-th/9408025
  [hep-th]}}.

\bibitem{Hlozek:2014lca}
R.~Hlozek, D.~Grin, D.~J.~E. Marsh, and P.~G. Ferreira, ``{A search for
  ultralight axions using precision cosmological data},''
  \href{http://dx.doi.org/10.1103/PhysRevD.91.103512}{{\em Phys. Rev.} {\bf
  D91} (2015) no.~10, 103512},
\href{http://arxiv.org/abs/1410.2896}{{\tt arXiv:1410.2896 [astro-ph.CO]}}.

\bibitem{D'Amico:2016kqm}
G.~D'Amico, T.~Hamill, and N.~Kaloper, ``{Quantum Field Theory of Interacting
  Dark Matter/Dark Energy: Dark Monodromies},''
\href{http://arxiv.org/abs/1605.00996}{{\tt arXiv:1605.00996 [hep-th]}}.

\bibitem{Unnikrishnan:2010ag}
S.~Unnikrishnan and L.~Sriramkumar, ``{A note on perfect scalar fields},''
  \href{http://dx.doi.org/10.1103/PhysRevD.81.103511}{{\em Phys. Rev.} {\bf
  D81} (2010)  103511},
\href{http://arxiv.org/abs/1002.0820}{{\tt arXiv:1002.0820 [astro-ph.CO]}}.

\bibitem{Arroja:2010wy}
F.~Arroja and M.~Sasaki, ``{A note on the equivalence of a barotropic perfect
  fluid with a K-essence scalar field},''
  \href{http://dx.doi.org/10.1103/PhysRevD.81.107301}{{\em Phys. Rev.} {\bf
  D81} (2010)  107301},
\href{http://arxiv.org/abs/1002.1376}{{\tt arXiv:1002.1376 [astro-ph.CO]}}.

\bibitem{Akhoury:2008nn}
R.~Akhoury, C.~S. Gauthier, and A.~Vikman, ``{Stationary Configurations Imply
  Shift Symmetry: No Bondi Accretion for Quintessence / k-Essence},''
  \href{http://dx.doi.org/10.1088/1126-6708/2009/03/082}{{\em JHEP} {\bf 03}
  (2009)  082},
\href{http://arxiv.org/abs/0811.1620}{{\tt arXiv:0811.1620 [astro-ph]}}.

\bibitem{Burgess:2007pt}
C.~P. Burgess, ``{Introduction to Effective Field Theory},''
  \href{http://dx.doi.org/10.1146/annurev.nucl.56.080805.140508}{{\em Ann. Rev.
  Nucl. Part. Sci.} {\bf 57} (2007)  329--362},
\href{http://arxiv.org/abs/hep-th/0701053}{{\tt arXiv:hep-th/0701053
  [hep-th]}}.

\bibitem{Goon:2016ihr}
G.~Goon, K.~Hinterbichler, A.~Joyce, and M.~Trodden, ``{Aspects of Galileon
  Non-Renormalization},''
\href{http://arxiv.org/abs/1606.02295}{{\tt arXiv:1606.02295 [hep-th]}}.

\bibitem{deRham:2014wfa}
C.~de~Rham and R.~H. Ribeiro, ``{Riding on irrelevant operators},''
  \href{http://dx.doi.org/10.1088/1475-7516/2014/11/016}{{\em JCAP} {\bf 1411}
  (2014) no.~11, 016},
\href{http://arxiv.org/abs/1405.5213}{{\tt arXiv:1405.5213 [hep-th]}}.

\bibitem{Felder:2002sv}
G.~N. Felder, L.~Kofman, and A.~Starobinsky, ``{Caustics in tachyon matter and
  other Born-Infeld scalars},''
  \href{http://dx.doi.org/10.1088/1126-6708/2002/09/026}{{\em JHEP} {\bf 09}
  (2002)  026},
\href{http://arxiv.org/abs/hep-th/0208019}{{\tt arXiv:hep-th/0208019
  [hep-th]}}.

\bibitem{deRham:2016ged}
C.~de~Rham and H.~Motohashi, ``{Caustics for Spherical Waves},''
\href{http://arxiv.org/abs/1611.05038}{{\tt arXiv:1611.05038 [hep-th]}}.

\bibitem{Mukohyama:2016ipl}
S.~Mukohyama, R.~Namba, and Y.~Watanabe, ``{Is the DBI scalar field as fragile
  as other $k$-essence fields?},''
  \href{http://dx.doi.org/10.1103/PhysRevD.94.023514}{{\em Phys. Rev.} {\bf
  D94} (2016) no.~2, 023514},
\href{http://arxiv.org/abs/1605.06418}{{\tt arXiv:1605.06418 [hep-th]}}.

\bibitem{Babichev:2016hys}
E.~Babichev, ``{Formation of caustics in k-essence and Horndeski theory},''
  \href{http://dx.doi.org/10.1007/JHEP04(2016)129}{{\em JHEP} {\bf 04} (2016)
  129},
\href{http://arxiv.org/abs/1602.00735}{{\tt arXiv:1602.00735 [hep-th]}}.

\bibitem{Babichev:2017lrx}
E.~Babichev and S.~Ramazanov, ``{Caustic free completion of pressureless
  perfect fluid and k-essence},''
\href{http://arxiv.org/abs/1704.03367}{{\tt arXiv:1704.03367 [hep-th]}}.

\bibitem{Frolov:2002rr}
A.~V. Frolov, L.~Kofman, and A.~A. Starobinsky, ``{Prospects and problems of
  tachyon matter cosmology},''
  \href{http://dx.doi.org/10.1016/S0370-2693(02)02582-0}{{\em Phys. Lett.} {\bf
  B545} (2002)  8--16},
\href{http://arxiv.org/abs/hep-th/0204187}{{\tt arXiv:hep-th/0204187
  [hep-th]}}.

\end{thebibliography}\endgroup

\end{document}